\documentclass[a4paper]{iopart}

\usepackage{iopams}     
\usepackage{graphicx}
\usepackage{epsfig}
\usepackage{color}
\usepackage{url}
\usepackage{times}
\usepackage{bm}         
\usepackage{times}

\newcommand{\bea}{\begin{eqnarray}}
\newcommand{\eea}{\end{eqnarray}}
\newcommand{\beq}{\begin{equation}}
\newcommand{\eeq}{\end{equation}}

\newcommand{\cf}{\textit{cf.~}}
\newcommand{\ie}{\textit{i.e.,}~}
\newcommand{\eg}{\textit{e.g.,}~}
\newcommand{\msec}{{\rm ms}}
\newcommand{\km}{{\rm km}}
\newcommand{\m}{{\rm m}}

\newcommand{\khz}{{\rm kHz}}
\newcommand{\mpc}{{\rm Mpc}}

\begin{document}

\title[Evolutions of inspiralling neutron-star binaries: assessment of
  the truncation error]{Accurate evolutions of inspiralling
  neutron-star binaries: assessment of the truncation error}

\author{Luca Baiotti$^{1}$, Bruno Giacomazzo$^{2}$, and Luciano Rezzolla$^{2,3}$}

\address{$^{1}$ Graduate School of Arts and Sciences, University of
  Tokyo, Tokyo, Japan }

\address{$^{2}$ Max-Planck-Institut f\"ur Gravitationsphysik,
  Albert-Einstein-Institut, Potsdam, Germany }

\address{$^{3}$ Department of Physics and Astronomy, Louisiana State
  University, Baton Rouge, LA, USA }

\date{\today}

\begin{abstract}
   We have recently presented an investigation in full
   general relativity of the dynamics and gravitational-wave emission
   from binary neutron stars which inspiral and merge, producing a
   black hole surrounded by a torus~\cite{Baiotti08}. We here discuss
   in more detail the convergence properties of the results presented
   in~\cite{Baiotti08} and, in particular, the deterioration of the
   convergence rate at the merger and during the survival of the
   merged object, when strong shocks are formed and turbulence
   develops. We also show that physically reasonable and numerically
   convergent results obtained at low-resolution suffer however from
   large truncation errors and hence are of little physical use. We
   summarize our findings in an \textit{``error budget''}, which
   includes the different sources of possible inaccuracies we have
   investigated and provides a first quantitative assessment of the
   precision in the modelling of compact fluid binaries.
\end{abstract}



\section{Introduction}
\label{sec:introduction}

The study of the final stages of the evolution of binary systems is a
cornerstone of any theory of gravity and a long-standing problem in
General Relativity. Important issues in relativistic astrophysics
still awaiting clarification, such as the mechanism responsible for
gamma-ray bursts (GRBs), may be unveiled through a better understanding
of the complex physics accompanying the inspiral and merger of two
neutron stars. Furthermore, the study of the events that lead from a
binary system of neutron stars to a black hole, possibly surrounded by
a hot and high-density disc, will provide the gravitational waveforms
and the energetics of one of the most important sources of
gravitational radiation. While analytical techniques are very
successful in describing binary systems which are widely separated and
thus moving at comparatively small velocities and in slowly varying
gravitational fields, numerical simulations represent possibly the
only tool to investigate the dynamics when the two compact objects are
performing the final few orbits of their evolution and the
dynamics are in a fully nonlinear regime.

Considerable progress has been achieved in the past few years in the
modelling in full general relativity of compact-objects binaries
(see~\cite{Anderson2007,Anderson2008,Duez:2008rb,Yamamoto2008,Etienne08,Etienne:2008re,Rezzolla:2007rd,Baker:2008,Campanelli:2008,Gonzalez:2008,Scheel:2008}
for the most recent work from the different groups), and we have
recently presented a systematic investigation of the dynamics and
gravitational-wave emission from binary neutron stars which inspiral
and merge, producing a black hole surrounded by a
torus~\cite{Baiotti08} (hereafter paper I). The purpose of this paper
is to consider in more detail the results presented in paper I and
assess critically their accuracy, their convergence properties and
highlight the difficulties that can be encountered when simulating the
turbulent motions that develop after the merger and that characterize
the subsequent evolution of the merged object up to the collapse to a
black hole.

The plan of the paper is as follows. After a brief introduction in
Section~\ref{sec:NumericalSpecifications} to our code and to the binary
models we simulated, we give in Section~\ref{sec:doc} a review of the
general dynamics of the inspiral and merger of equal-mass neutron star
binaries. Section~\ref{sec:conv-acc} is instead dedicated to the
discussion of the accuracy of our simulations when performed with
different resolutions, while Section~\ref{sec:errors} assesses the
influence that purely numerical aspects have on the physical results
and presents them in a compact \textit{``error budget''}. Finally, our
summary and conclusions are collected in Section~\ref{sec:conclusion}.


\section{Mathematical and Numerical Setup}
\label{sec:NumericalSpecifications}

All the details on the mathematical and numerical setup used for
producing the results presented here are discussed in depth
in~\cite{Pollney:2007ss,Baiotti08}. In what follows, we limit
ourselves to a brief overview.


\subsection{Einstein and Hydrodynamics equations}
\label{sec:Einsten_Hydro_eqs}

The evolution of the spacetime was
obtained using the \texttt{Ccatie} code, a three-dimensional
finite-differencing code providing a solution of a conformal
traceless formulation of the Einstein
equations~\cite{Pollney:2007ss}. The general-relativistic equations
were instead solved using the \texttt{Whisky} code presented
in~\cite{Baiotti03a,Baiotti04,Baiotti07}, which adopts a flux-conservative formulation of
the equations as presented in~\cite{Banyuls97} and high-resolution
shock-capturing schemes. The {\tt Whisky} code implements several
reconstruction methods, such as Total-Variation-Diminishing (TVD)
methods, Essentially-Non-Oscillatory (ENO) methods~\cite{Harten87} and
the Piecewise Parabolic Method (PPM)~\cite{Colella84}. Also, a variety
of approximate Riemann solvers can be used, starting from the
Harten-Lax-van Leer-Einfeldt (HLLE) solver~\cite{Harten83}, over to
the Roe solver~\cite{Roe81} and the Marquina flux
formula~\cite{Aloy99b} (see~\cite{Baiotti03a,Baiotti04} for a more
detailed discussion). All the results reported hereafter have been
computed using the Marquina flux formula and a PPM reconstruction.  We
stress again (as already done in~\cite{Baiotti08,Giacomazzo:2009mp})
that the use of high-order methods and high-resolution is
\textit{essential} to be able to draw robust conclusions on the
inspiral and merger. Lower-order methods in the reconstruction and low
resolution may yield convergent and apparently reasonable results
which however contain a large truncation error. Specific examples of
this type of problem are presented in Appendix 1 of paper I and in
Figure 4 of~\cite{Giacomazzo:2009mp}.

The system of hydrodynamics equations is closed by an equation of
state (EOS) and, as discussed in detail in~\cite{Baiotti08}, the
choice of the EOS plays a fundamental role in the post-merger dynamics
and significantly influences the survival time, against gravitational
collapse, of the hyper-massive neutron star (HMNS) likely produced by
the merger. It is therefore important that special attention is paid
to use EOSs that are physically realistic, as done
in~\cite{Oechslin07b} within a conformally flat description of the fields
and a simplified treatment of the hydrodynamics. Because we are here
mostly concerned with assessing the size of the truncation error
rather than with a realistic description of the neutron-star matter,
we have employed the commonly used ``ideal-fluid'' EOS, in which the
pressure $p$ is expressed as $p = \rho\, \epsilon(\Gamma-1) $, where
$\rho$ is the rest-mass density, $\epsilon$ is the specific internal
energy and $\Gamma$ is the adiabatic exponent. Such an EOS, while simple,
provides a reasonable approximation and we expect that the use of
realistic EOSs would not change the main results of this work.


\subsection{Adaptive Mesh Refinements}
\label{sec:AMR}

Both the Einstein and the hydrodynamics equations are solved using the
vertex-centered adaptive mesh-refinement (AMR) approach provided by
the \texttt{Carpet} driver~\cite{Schnetter-etal-03b}. Our rather basic 
form of AMR consists in centering the highest-resolution level around
the peak in the rest-mass density of each star and in moving the ``boxes''
so to track the position of this maximum as the stars orbit. The boxes are 
merged when they overlap.

The results presented below refer to simulations performed at three
different resolutions and for each of them we have used $6$ levels of
mesh refinement. More specifically, the finest refinement level has
been chosen to have resolution of either $h=0.1875\,M_{\odot}=277\,\m$
(hereafter ``low'' resolution), of $h=0.1500\,M_{\odot}=222\,\m$
(hereafter ``medium'' resolution), or of $h=0.1200\,M_{\odot}=177\,\m$
(hereafter ``high'' resolution). It may be useful to point out that
although the simulation at high-resolution stretches the computational
resources available to us, it is effectively more expensive (both in
memory and computing time) than an equivalent simulation carried out
for binary black holes at resolutions that are $10$ times larger but
on fine grids that are much smaller (\cf the high-resolution
in~\cite{Pollney:2007ss}). Using the medium resolution as a reference,
the grid structure is such that the side of the finest grids is
$15\,M_{\odot}=22.15\,\km$, while a single grid resolution covers the
region between a distance $r=150\,M_{\odot}=221.5\,\km$ and
$r=250\,M_{\odot}=369.2\,\km$ from the center of the domain. The
latter region is the one in which our gravitational-wave extraction is
carried out, with a resolution of $h=4.8\,M_{\odot}=7.088\,\km$ (as a
comparison, the gravitational wavelength is $\sim 100\,\km$ and thus
reasonably well-resolved on this grid). In addition, a set of refined
but fixed grids is set up at the center of the computational domain so
as to capture the details of the Kelvin-Helmholtz instability (\cf
paper I). Unless explicitly stated, for all the simulations reported here we have used a
reflection-symmetry condition across the $z=0$ plane and a
$\pi$-symmetry condition across the $x=0$ plane.

An important difference with respect to paper I, where the finest grid
was covering only the central region of each neutron star, is that
here each star is completely covered by the finest grid. Although this
choice is computationally more expensive, it allows us to reach
convergent results already with resolutions $h\gtrsim
0.19\,M_{\odot}$, which are therefore smaller than those discussed in
paper I (where we have used what is here the high resolution).

The timestep on each grid is set by the Courant condition (expressed
in terms of the speed of light) and so by the spatial grid resolution
for that level; the typical Courant coefficient is set to be $0.35$.
The time evolution is carried out using $4$th-order--accurate
Runge-Kutta integration algorithm. Boundary data for finer grids are
calculated with spatial prolongation operators employing $3$rd-order
polynomials and with prolongation in time employing $2$nd-order
polynomials. The latter allows a significant memory saving, requiring
only three timelevels to be stored, with little loss of accuracy due
to the long dynamical timescale relative to the typical grid timestep.
See Section~\ref{sec:errors} for a discussion on the changes caused
by the different interpolation order.


\subsection{Initial data}
\label{sec:initial_data}

As initial data we use the general relativistic binaries produced by
Taniguchi and Gourgoulhon~\cite{Taniguchi02b} with the multidomain
spectral-method code {\tt Lorene}~\cite{lorene}. The initial solutions
for the binaries are obtained assuming a quasi-circular orbit, an
irrotational velocity field, and a conformally-flat spatial
metric. The matter is modelled using a polytropic EOS $p = K
\rho^{\Gamma}$ with $K=123.6$ and $\Gamma=2$.

In paper I, a number of different initial-data configurations for
neutron-star binaries were used in order to illustrate the variety of
possible behaviours. Here, however, the measure of the truncation
error is more easily done when considering a single configuration and
in particular one that, although representative, also reduces the
computational costs. As a result, we have chosen a model with a rather
high mass so that a black hole is formed soon after the merger. More
specifically, our fiducial binary has the following characteristics: a
proper separation between the centers of the stars $d/M_{_{\rm
    ADM}}=12.6=60.3\,\km$ (this corresponds to a coordinate distance
of $\sim 45\,\km$); a baryon mass of each star $M_{b}=1.78\,M_{\odot}$;
a total ADM mass $M_{_{\rm ADM}}=3.23\,M_{\odot}$; an initial angular
momentum $J=10.13\,M^2_{\odot}= 8.92\times 10^{49}\,{\rm g\,cm^2/s}$;
an initial orbital angular velocity $\Omega_0=9.39\times10^{-3}= 1.9\,
{\rm rad/ms}=302\,{\rm Hz}$; a ratio of the polar to the equatorial
coordinate radius of each star $r_p/r_e=0.945$.

A note of caution should be made. While our choice for a rather
massive binary does have the advantage of allowing, within a total
timescale of $\sim 11\,\msec$, for the analysis of \textit{both} the
inspiral (over more than three orbits) and of the merger, it also
leads to a very rapid collapse to a black hole once the HMNS has been
produced (\ie after only $\sim 1\,\msec$). This is simply because
despite the high temperature and degree of differential rotation, the
HMNS is so massive to be well beyond the instability threshold for the
collapse to a black hole. As a result, this choice inevitably prevents
us from determining how different choices of grids or resolutions
influence the survival time of the HMNS when it is very close to
the instability threshold. This is clearly a limitation of the present
approach which will be overcome once larger computational facilities
become available.

\section{Review of the dynamics of the merger}
\label{sec:doc}

In paper I we have described in detail the dynamics of the matter
during the inspiral, the merger, the transition to the collapse, the
collapse, and then the black-hole ringdown, for binaries with
different initial masses, initial distances, and (idealized) EOSs. In
what follows, we briefly summarize those results.

A first notable result of paper I is that of having clearly shown that
for any given mass, the survival time of the HMNS depends on the
EOS. More specifically, we have shown that the polytropic and
isentropic EOS $p=K\rho^{\Gamma}$, for which therefore no shock
heating is possible, leads either to the \textit{prompt} formation of
a rapidly rotating black hole surrounded by a dense torus in the
higher-mass case, or, in the lower-mass case, to a
HMNS, which develops a bar, emits large amounts of
gravitational radiation and eventually experiences a \textit{delayed}
collapse to black hole. We have also shown that, for both initial
masses, the ideal-fluid EOS inevitably leads to a further delay in the
collapse to black hole as a result of the larger pressure support
provided by the temperature increase due to shocks. In this case the
temperature in the formed HMNS can reach values as high as
$10^{11}-10^{12}\,{\rm K}$, so that the subsequent dynamics and
especially the time of the collapse can be reduced if cooling
mechanisms, such as the direct URCA\footnote{Direct URCA processes are processes that produce
  neutrinos, via the decay of a baryon or an interaction of a baryon with a lepton. The word URCA is
not an acronym, but the name of the place where these processes were first discussed by G. Gamow and
M. Schenberg.} process, take place.

With the exception of the low-mass ideal-fluid binary, whose HMNS is
expected to collapse to black hole on a timescale which is
computationally challenging (\ie $\sim 110\,\msec$), all the binaries
considered lead to the formation of a rotating black hole surrounded
by a rapidly rotating torus. The masses and dimensions of the tori
depend on the EOS, but are generically larger than those reported in
previous independent studies, with masses up to $\approx
0.07M_{\odot}$. Confirming what was reported in~\cite{Shibata06a}, we
have found that the amount of angular momentum lost during the
inspiral phase can influence the mass of the torus for binaries that
have the same EOS. In particular, the binaries that lose less angular
momentum during the inspiral, namely the comparatively
\textit{low-mass} binaries, are expected to have comparatively
\textit{high-mass} tori. In addition, we have also considered the
comparison of binaries of the same mass but with different initial
coordinate separation (\ie $45$ and $60\,\km$) and found that there is
an excellent agreement in the inspiral phase (as expected from the
lowest-order post-Newtonian approximations), but also small
differences at the merger and in the subsequent evolution, most likely
due to small differences in the initial data.

Besides the study of the large-scale dynamics of the two neutron
stars, in paper I we have also investigated the small-scale
hydrodynamics of the merger and the possibility that dynamical
instabilities develop. In this way, we have provided the first
quantitative description of the onset and development of the
Kelvin-Helmholtz instability, which takes place during the first
stages of the merger phase, when the outer layers of the stars come
into contact and a shear interface forms. The instability curls the
interface forming a series of vortices which we were able to resolve
accurately using the higher resolutions provided by AMR
techniques. This instability, which could have important consequences
in the generation of large magnetic fields even from small initial
ones, has been recently discussed in~\cite{Giacomazzo:2009mp}.

Special attention in paper I was also dedicated to the analysis of the
waveforms produced and to their properties for the different
configurations. In particular, we have found that the largest loss
rates of energy and angular momentum via gravitational radiation
develop at the time of the collapse to black hole and during the first
stages of the subsequent ringdown. Nevertheless, the configurations
which emit the largest amounts of energy and angular momentum are
those with lower masses, since they do not collapse promptly to a
black hole. Instead, they produce a violently oscillating HMNS, which
emits copious gravitational radiation, while rearranging its
angular-momentum distribution, until the onset of the collapse to
black hole. We have also found that, although the gravitational-wave
emission from neutron-star binaries has spectral distributions with
large powers at high frequencies (\ie $\gtrsim 1\,\khz$), a
signal-to-noise ratio as large as $3$ can be estimated for a source at
$10\,\mpc$ using the sensitivity of currently operating
interferometric detectors.


\section{Accuracy of the results}
\label{sec:conv-acc}

\begin{figure}[t]
\begin{center}
   \includegraphics[width=0.49\textwidth]{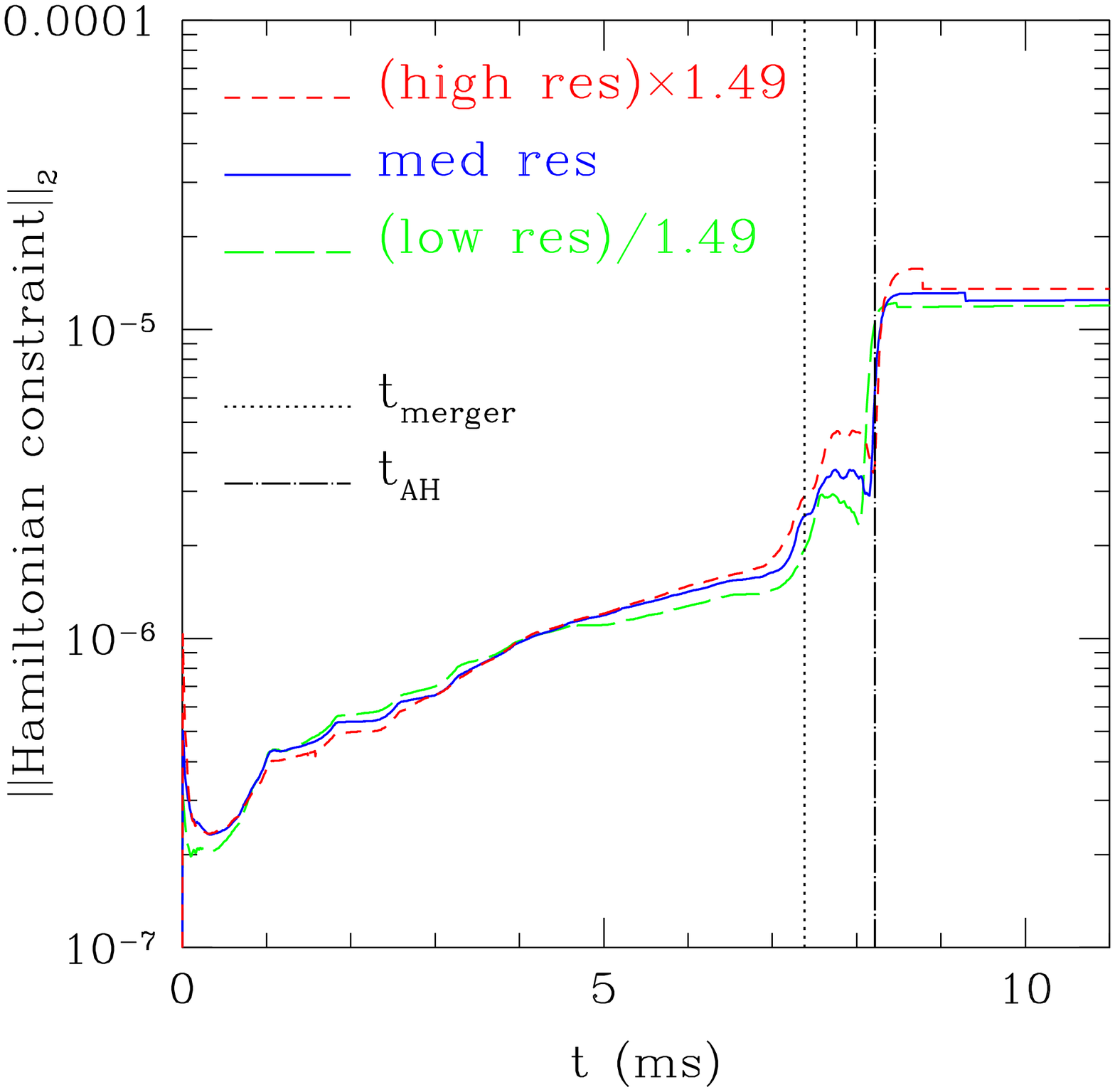}
   \includegraphics[width=0.49\textwidth]{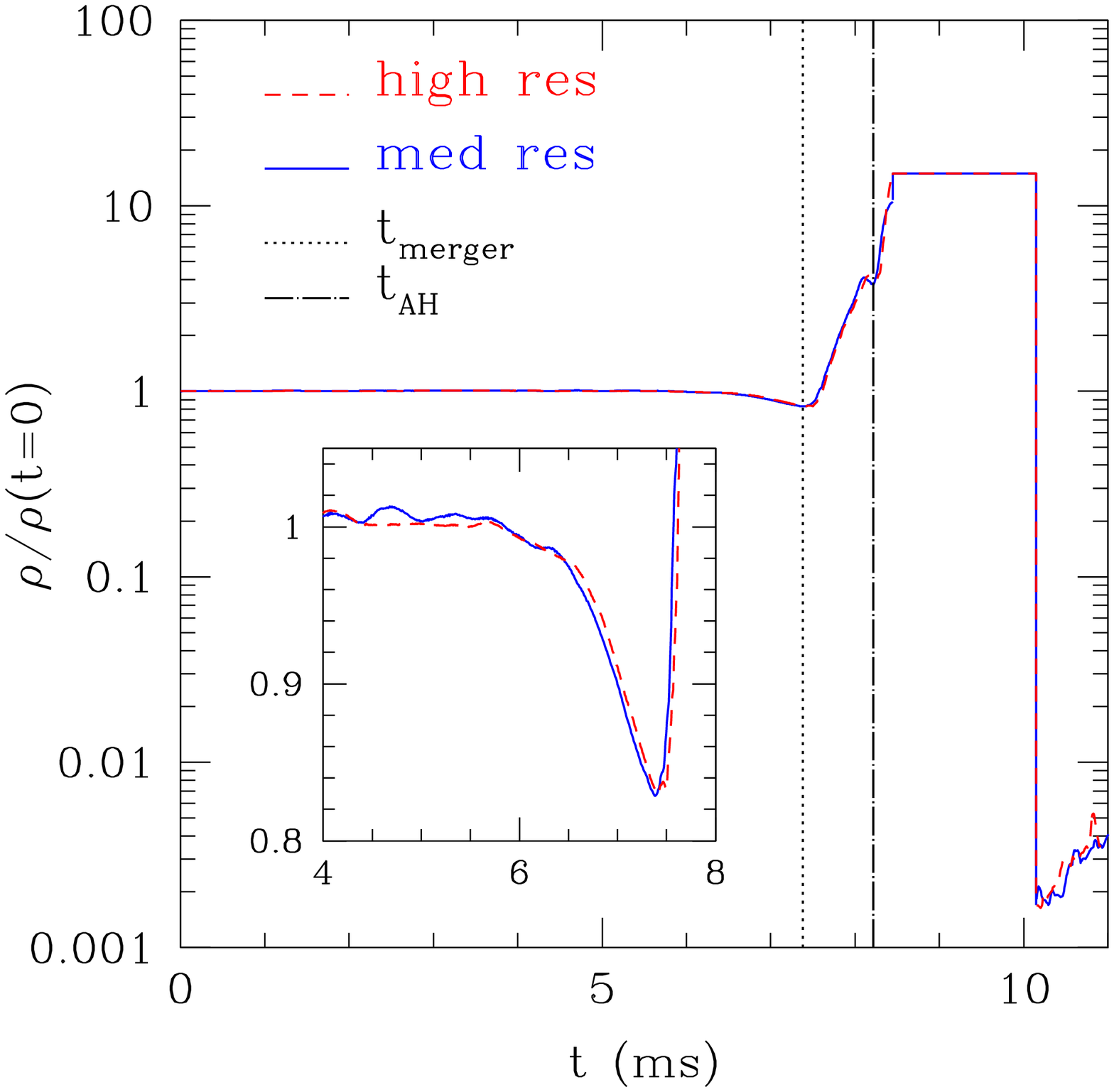}
\end{center}
\vskip -0.5cm
   \caption{\textit{Left panel:} Evolution of the $L_2$-norm of the
     Hamiltonian-constraint violation for simulations at high
     resolution (dashed line), at medium resolution (solid line) and
     at low resolution (long-dashed line). Note that the high- and
     low-resolution curves are rescaled by $1.49$, which corresponds
     to a convergence rate of $1.8$. The vertical lines indicate the
     time of merger $t_{\rm merger}$ for the high-resolution run
     (dotted line) and the time when an apparent horizon is first
     found $t_{_{\rm AH}}$ (dot-dashed line). \textit{Right panel:}
     Evolution of the maximum of the rest-mass density normalized at
     the initial value and for the high and medium resolutions. Note
     that the horizontal lines mark the part of the evolution when the
     rest-mass density is unreliable after the formation of the
     apparent horizon; when the matter inside the apparent horizon has
     been dissipated, the maximum of the rest-mass density is that of
     the torus (see text for details).}
     \label{fig:ham}
\end{figure}

Having reviewed the general dynamics of the inspiral and merger, we
next assess the truncation error of the results. As a first
representative measure of the accuracy of our simulations, we report
in the left panel of Figure~\ref{fig:ham} the evolution of the
$L_2$-norm of the Hamiltonian-constraint violation [\ie equation~(10)
  in paper I] for simulations at high resolution (dashed line), at
medium resolution (solid line) and at low resolution (long-dashed
line). Although the resolutions are rather similar (their ratio is
just $0.8$), they are high enough to stretch the supercomputer
facilities we have access to. Furthermore, as we will discuss below,
they are only marginally sufficient to provide convergent results and
resolutions lower than $h\simeq 0.19\,M_{\odot}$ would lead to results
that are only consistent\footnote{We recall that a numerical solution
  is said to be \textit{consistent} if the truncation error associated
  to it, $\epsilon_{_{\rm T}}$, tends to zero in the limit of infinite
  resolution, \ie $|\epsilon_{_{\rm T}}| = 0$ for $h \to 0$. A
  numerical solution is said to be \textit{convergent} at the order
  $p$ if and only if $|\epsilon_{_{\rm T}}| = \kappa h^p$ for $h\to
  0$, where $\kappa$ is a positive constant and $p$ is the
  \textit{chosen} truncation order. Clearly, while a convergent
  solution is also consistent, the opposite is not necessarily
  true.}. The curves relative to the high and low resolutions have
been scaled by a suitable factor compensating for the different
truncation error. The vertical lines are relative to the
high-resolution run and indicate respectively the time of the merger
$t_{\rm merg.}$ (dotted line) and the time when an apparent horizon is
first found $t_{_{\rm AH}}$ (dot-dashed line). Taking the $L_2$-norm
of the same constraint as a measure of the average truncation error,
Figure~\ref{fig:ham} shows that this is initially $\lesssim 10^{-6}$
and that it grows rapidly to $\lesssim 10^{-4}$ at the time of the
merger and later when the black hole is formed. As a comparison, the
$L_{\infty}$-norm of the violation of the Hamiltonian constraint grows
from $\sim 10^{-3}$ initially to $\sim 10^{-2}$ just outside the
apparent horizon; much larger violations, \ie $L_{\infty}\sim 1$, are
present inside the apparent horizon and are typical of these
calculations (\cf~\cite{Baiotti06}).

Once the black hole has essentially stopped ringing, the violation
does not increase further. The convergence rate measured before the
merger, thus involving $\sim 3$ orbits, is therefore $\simeq 1.8$ as
shown by the good overlap between the mid-resolution and the rescaled
high/low-resolution curves. We note that: although {\it i}) the
\texttt{Ccatie} code uses $4$th-order spatial differential operators,
       {\it ii}) the time update is made with a $4$th-order accurate
       Runge-Kutta integration scheme, and {\it iii}) the PPM
       reconstruction scheme is $3$rd-order (at most), a convergence
       rate of $\simeq 1.8$ is not at all surprising and is indeed the
       same measured with PPM in much simpler configurations, such as
       in the evolution of isolated stars~\cite{Baiotti04}\footnote{When 
         local maxima (\eg at a stellar centre) and/or small shocks in the 
         vicinity of the stellar surface (see the discussion in Sect. III\,C
         of~\cite{Baiotti08}) are present - that is, in the interval preceding the 
         collapse in the simulations reported here - the convergence order is lower 
         than third; fortunately these errors are small enough not to degrade the 
         global convergence order to first order only.}.
This is simply due to the fact that the
truncation error of HRSC schemes is not uniform across the
computational domain and can drop to lower orders at local
extrema or across shocks.

After the merger, the convergence rate drops to $\simeq 1.2$ and this
deterioration may be due to at least two different reasons and is
probably due to both. Firstly, we note that the merger is accompanied
by very large and extended shocks and, as mentioned above, a general
property of HRSC methods is to become only $1$st-order accurate at
discontinuities. Secondly, the merger leads to the development of a
Kelvin-Helmholtz instability with the consequent generation of a
turbulent regime (see the discussion in Section~III\,E of paper
I). Under these conditions, the whole notion of convergence needs to
be revisited and a possible approach in these cases is the one
presented in~\cite{Fromang2007}. Note also that the convergence order
increases again after the formation of the black hole, when only a
small amount of matter is present in the form of an orbiting torus.
Such flows do not have large-scale shocks, are much smoother and hence
lead to convergence rates that are again $\sim 1.8$. Clearly, tests at
even higher resolutions should be performed to determine the precise
cause of the degradation in the convergence order, but it is unlikely that
these will be feasible with the present computational resources. As a
final remark, it is worth reporting that a behaviour similar to the
one discussed here has been found~\cite{Baiotti09_private} also in the
independent AMR code {\tt SACRA}~\cite{Yamamoto2008} (based on similar
numerical methods).

\begin{figure}[h]
\begin{center}
   \includegraphics[width=0.49\textwidth]{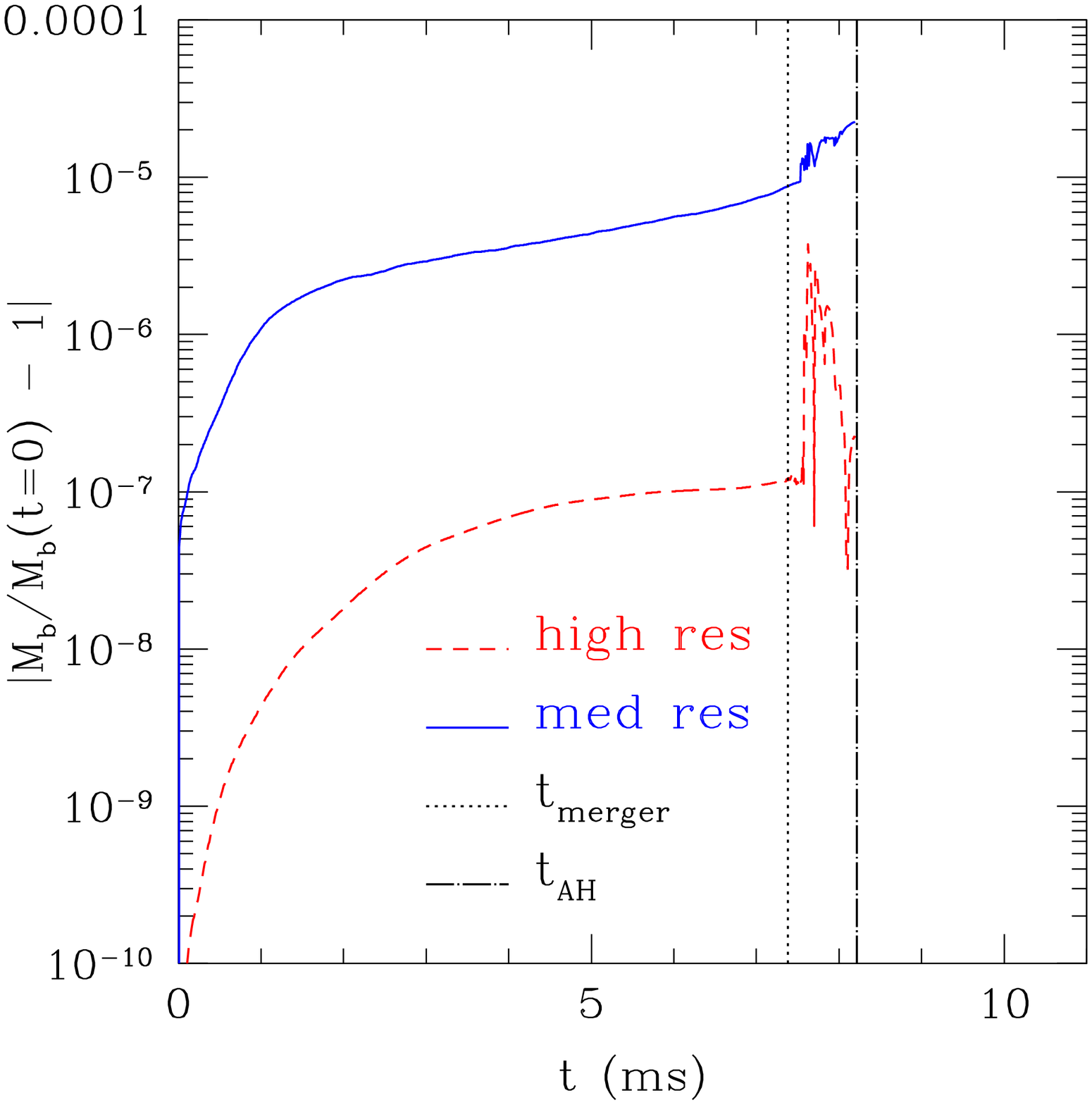}
   \includegraphics[width=0.49\textwidth]{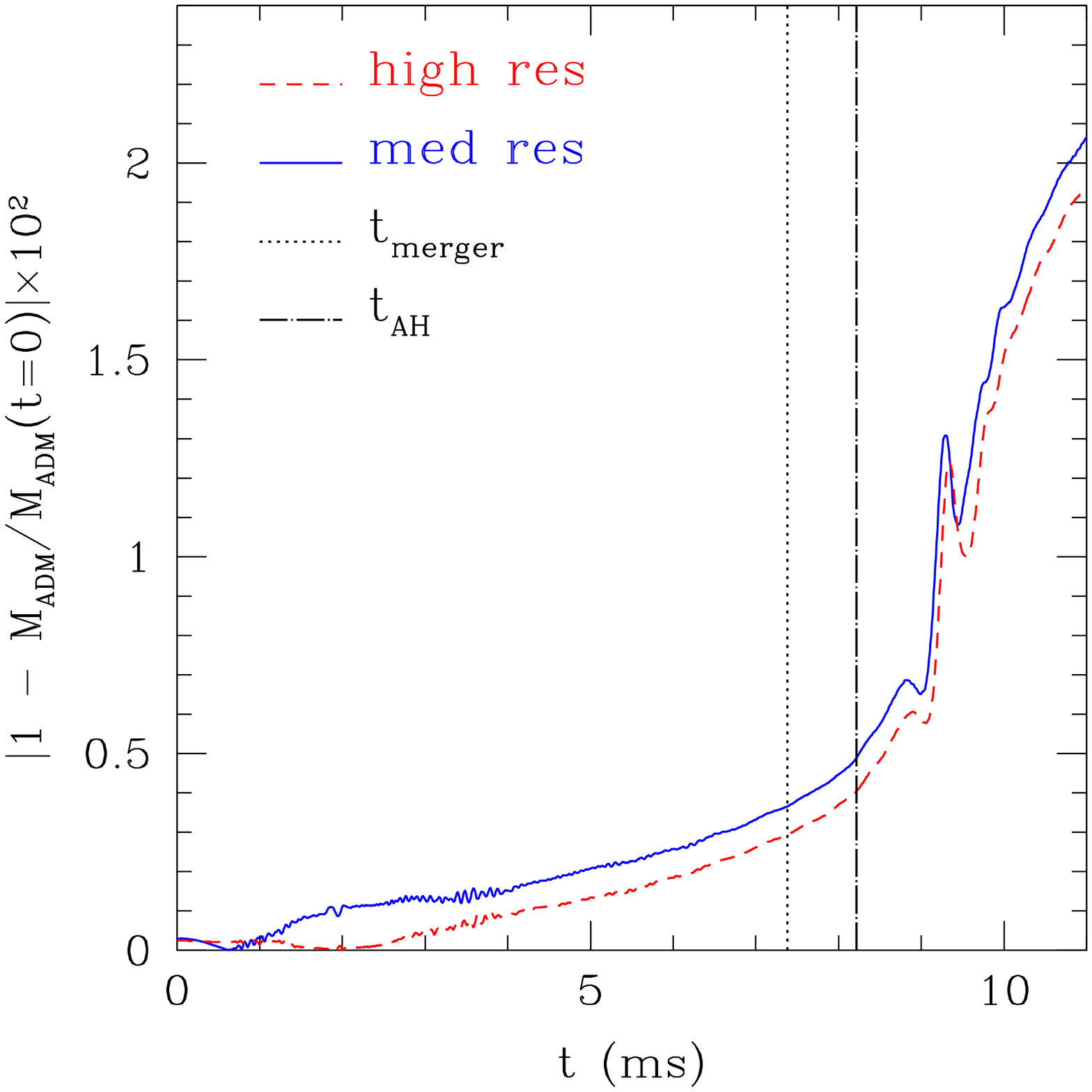}
\end{center}
\vskip -0.5cm
   \caption{\textit{Left panel:} Evolution of the error in the
     conservation of the rest mass $M_{b}$ for the medium and high
     resolutions. Note that the fluctuations in $M_{b}$ after the
     merger are due to the small amount of matter that ends up in
     buffer zones between two refinements levels. \textit{Right
       panel:} Evolution of the normalized ADM mass as measured on the
     numerical grid; until when significant amounts of gravitational
     radiation leave the numerical grid (\ie up until about $9$ ms),
     the differences between the two resolutions measure the error in
     computing $M_{_{\rm ADM}}$.}
     \label{fig:restmass}
\end{figure}

The right panel of Figure~\ref{fig:ham} shows instead the evolution of
the maximum of the rest-mass density when normalized to its initial
value and where, for clarity, we have reported only the medium and
high-resolution results (these will be the reference resolutions
hereafter). Clearly, the two lines show a behaviour which is extremely
similar during the inspiral phase, with only very small differences
appearing at the time of the merger and after the collapse. These
differences are caused mostly by the slight difference in the time of
the merger, which is $\simeq 1.7\%$ smaller in the case of the
lower-resolution simulation and which is probably due to the slightly
different initial data and losses of angular momentum. Note that we
have here used a technique, first reported in~\cite{Baiotti06}, where
we do not excise any part of the computational domain, but we rather
rely on suitable singularity-avoiding slicing conditions and on the
addition of a small amount of numerical dissipation to avoid that
small inaccuracies influence the portion of the spacetime exterior to
the apparent horizon. Since we do not reserve a special treatment to
the matter collapsing inside the apparent horizon, the matter
variables and their spatial gradients become very large. These are
reasonably well-handled by the HRSC schemes, but only up until all the
matter is confined to a few grid cells; at that point, the
conservation properties of our methods are grossly overstretched and
the matter is simply dissipated. Because of these inaccuracies inside
the apparent horizon, the evolution of the maximum of the rest-mass
density is unreliable after the formation of the apparent horizon and
up until the rest-mass inside the apparent horizon has become
negligible. This is shown with constant horizontal lines in the right
panel of Figure~\ref{fig:ham}. Note that when the matter inside the
apparent horizon has been dissipated, \ie for $t\gtrsim 10\,\msec$,
the maximum of the rest-mass density is located outside the apparent
horizon and thus represents the maximum rest-mass density of the
torus.

A first measure of the conservation properties of our simulations is
offered in Figure~\ref{fig:restmass}, whose left and right panels show
respectively the evolution of the error in the conservation of the
rest mass $M_{b}$ and of the ADM mass $M_{_{\rm ADM}}$. Note that
thanks to the formulation of the hydrodynamics equations adopted and
to the use of HRSC methods, the baryon mass is extremely well
conserved, with deviations that are $\sim 10^{-6}$ ($\sim 10^{-7}$ for
the high-resolution simulation) before the merger and are always below
$\sim 10^{-5}$ ($\sim 10^{-6}$) up to the formation of the black
hole. Note also that at the time of the merger the new grid structure
leads to a certain amount of matter ending up in buffer zones between
two refinements levels. This inevitably increases the error in the
conservation, which remains anyway extremely good.  This is
particularly evident for the high-resolution run, where the error is
intrinsically tiny. As discussed above, after the apparent horizon is
found the baryon mass is no longer conserved up until all the matter
inside the apparent horizon has been dissipated numerically. When this
has happened, the baryon mass is the one contained in the torus and is
equally well conserved, although it cannot be shown in the same panel
because much smaller.

The ADM mass cannot be conserved when calculated on the finite-size
numerical grid and it is rather expected to decrease as gravitational
radiation leaves the numerical grid. The right panel of
Figure~\ref{fig:restmass} shows the evolution of the ADM mass of the 
two reference resolutions when normalized to their initial value. 
Although this quantity should not be interpreted as an
error, the difference between the two resolutions gives a measure of
the truncation error in computing $M_{_{\rm ADM}}$. Such a difference
is $\sim 0.1\%$ for essentially all the simulations and the large
variations at $\sim 9\,\msec$ take place when the gravitational
radiation produced by the collapse to black hole leaves the
computational domain.

\begin{figure}[b]
\begin{center}
   \includegraphics[width=0.49\textwidth]{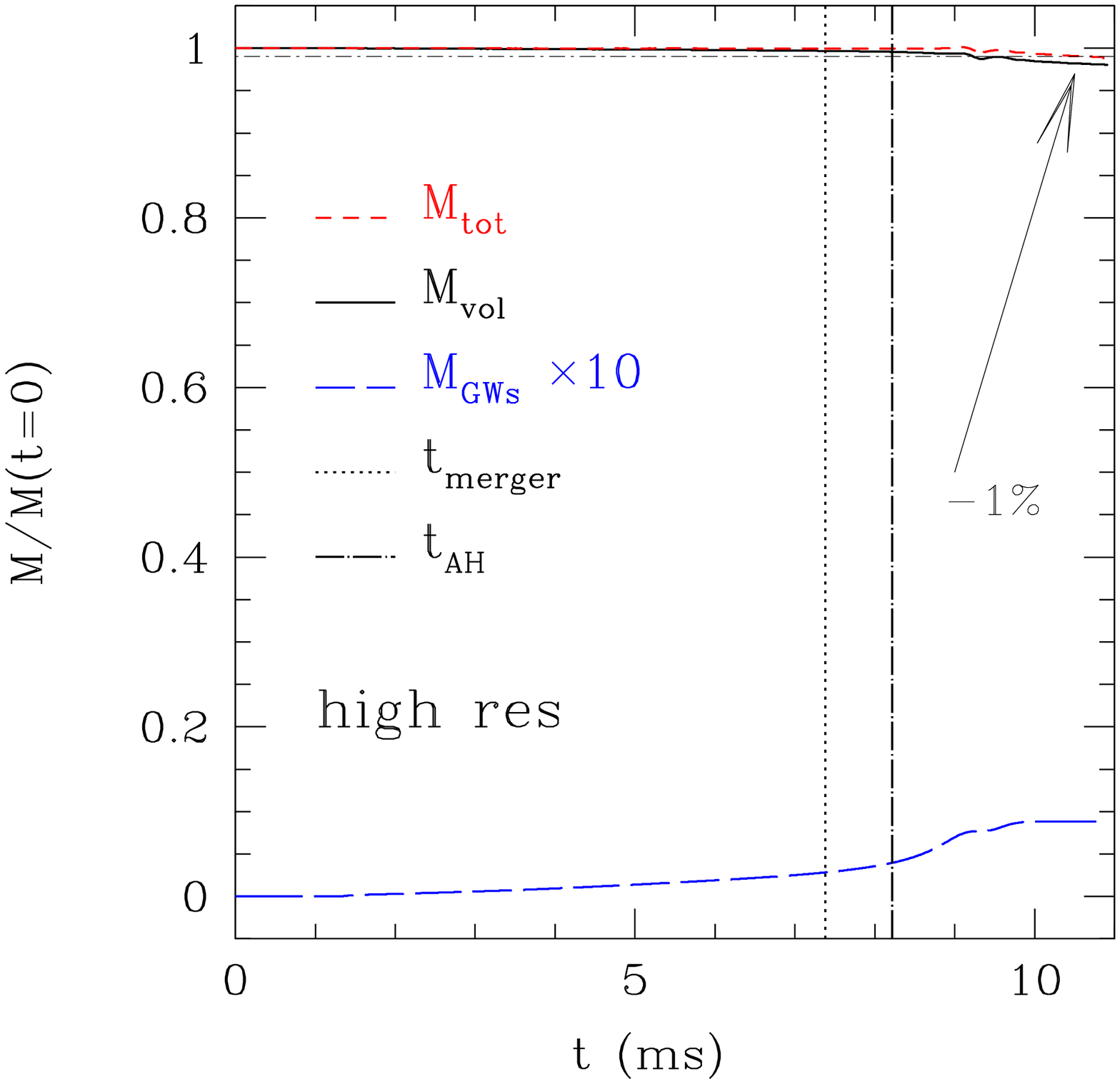}
   \includegraphics[width=0.49\textwidth]{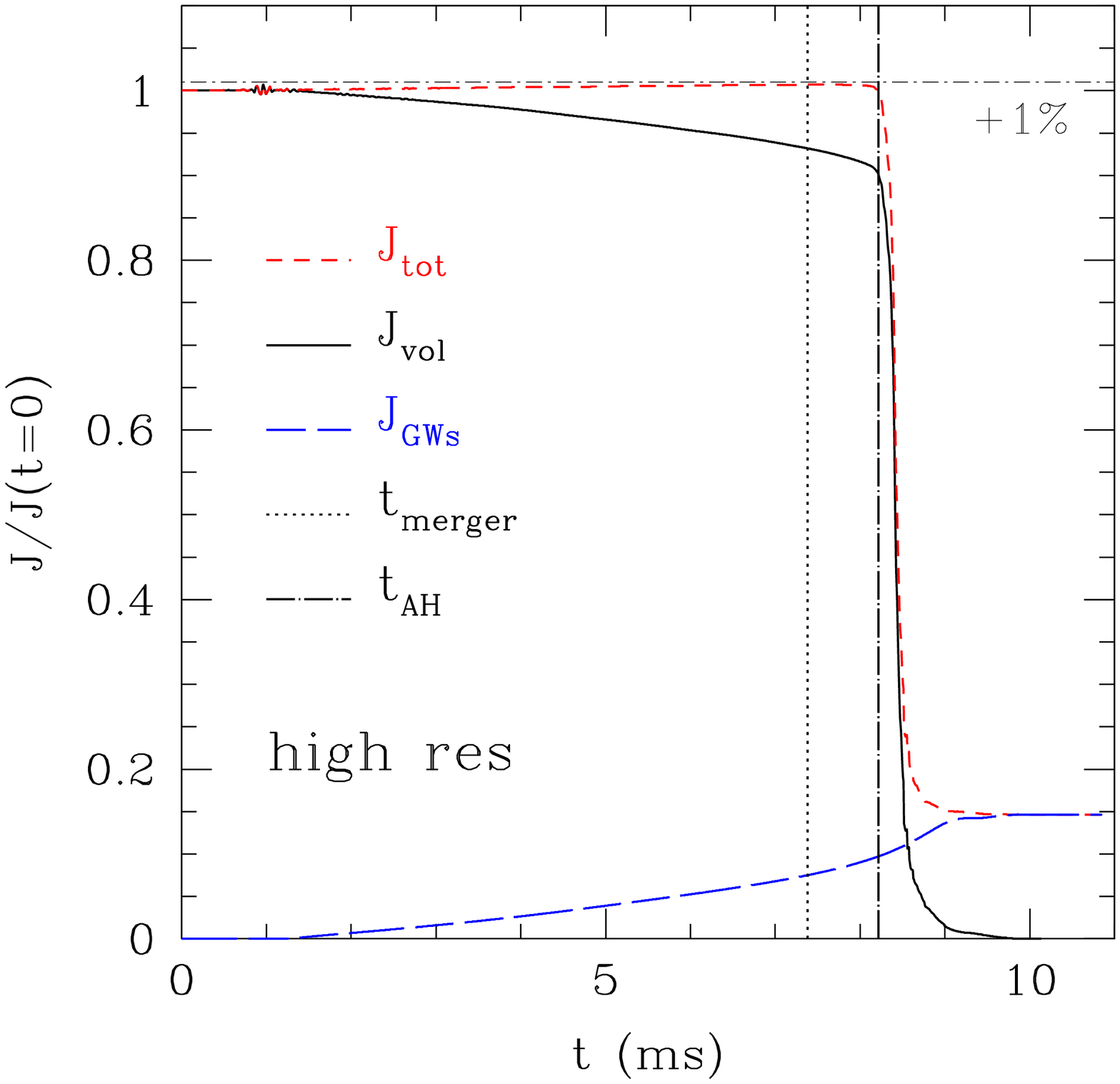}
\end{center}
\vskip -0.5cm
   \caption{\textit{Left panel:} conservation of energy. The
     continuous line is the ADM mass computed as an integral over the
     whole grid; The long-dashed line is the energy carried from
     gravitational waves outside the grid; The dashed line is the sum
     of the two and it should be conserved. The numerical violation is
     at most $1\%$. The data refer to the higher-resolution
     simulation.  \textit{Right panel:} the same as in the left panel
     but for the conservation of the angular momentum. Also in this
     case the violation is at most $1\%$.}
     \label{fig:am}
\end{figure}

A second and more stringent measure of the overall conservation
properties of our simulations is presented in Figure~\ref{fig:am} and
involves quantities which are partially radiated during the
simulation. More specifically, the left panel shows the evolution of
the total mass as normalized to the initial value and relative to the
high-resolution simulation. Indicated with different lines are
the volume-integrated values of the ADM mass (solid
line), of the energy lost to gravitational waves (long-dashed line), and of
their sum (dashed line). The last quantity should be strictly constant
and this is the case to a precision of $\sim 0.5\%$ during the
inspiral, but with a secular decreases that brings the total error to
be $\sim 1\%$ at the end of the simulation (\cf~dot-dashed
line). Similar considerations apply also to the conservation of the
angular momentum (\cf~Figure 9 of paper I). This is shown in the right
panel of Figure~\ref{fig:am}, which uses the same conventions as the
left panel [here $J_{\rm vol}$ is computed with the integral (15) in
  paper I]. In this case the radiative losses are much larger (almost
$15\%$ of the available angular momentum is lost to gravitational
waves) but, as shown in the figure, the overall conservation is
accurate to $\sim 1\%$.

\begin{figure}[t]
\begin{center}
   \includegraphics[width=0.49\textwidth]{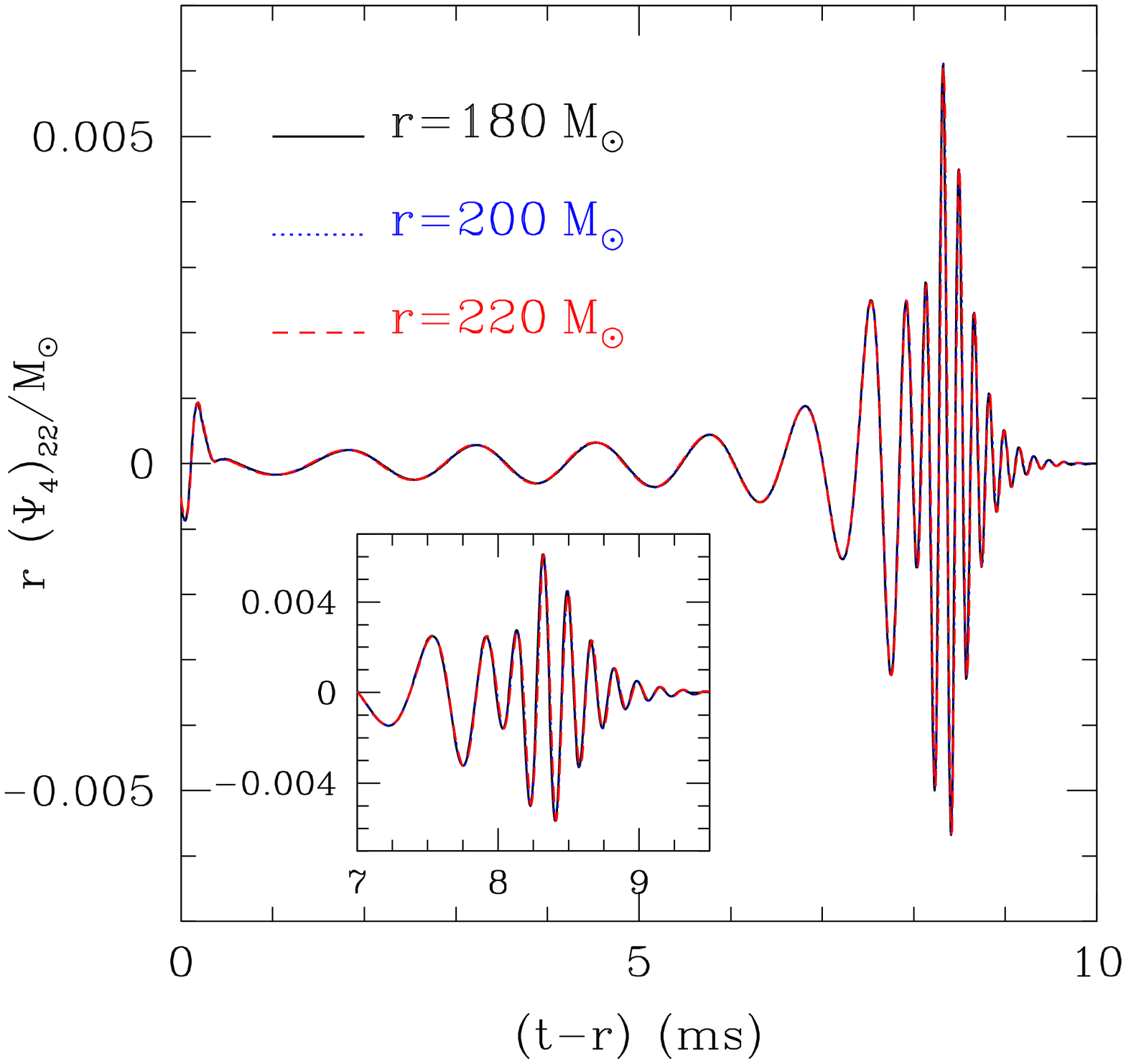}
   \includegraphics[width=0.49\textwidth]{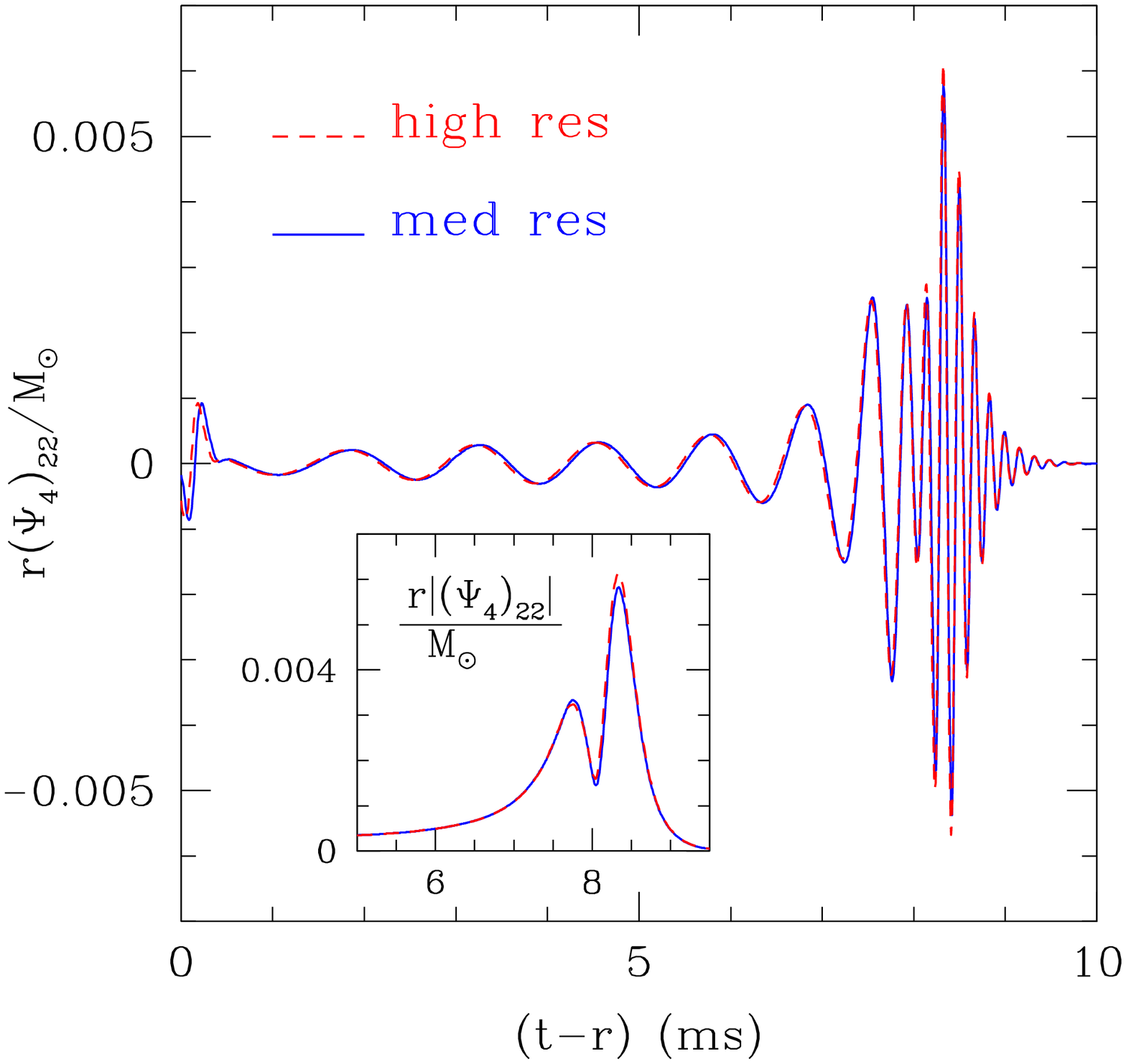}
\end{center}
\vskip -0.5cm
   \caption{\textit{Left panel:} $r(\Psi_4)_{22}/M_{\odot}$ waveform for
     different detectors, displayed versus retarded time. The data
     refer to the higher-resolution simulation. \textit{Right panel:}
     $r(\Psi_4)_{22}/M_{\odot}$ waveform as computed for different resolutions.}
     \label{fig:waves}
\end{figure}

The accuracy of our gravitational-wave extraction is summarized in
Figure~\ref{fig:waves}, where in the left panel we show the
gravitational-wave signal as computed via the $\ell=2,\,m=2$ component
of the Weyl scalar quantity $\Psi_4$, \ie $r(\Psi_4)_{22}/M_{\odot}$,
as a function of the retarded time. The data is that of the
high-resolution simulation and different lines refer to different
extraction $2$-spheres; the agreement among the three curves is
extremely good, with the phases and amplitudes being almost coincident
(see also the small inset). The right panel of Figure~\ref{fig:waves}
shows instead $r(\Psi_4)_{22}/M_{\odot}$ at the high and medium
resolutions, as extracted at $r=200\,M_{\odot}$. As mentioned above,
the slightly different initial data and losses of angular momentum
will lead to a slight offset of the waveforms as the inspiral proceeds
(this is of $\sim 0.03\,\msec$) and the data is shown once this offset
has been removed, \ie after aligning the maxima of
$|(\Psi_4)_{22}|$. Clearly the two waveforms are extremely similar and
small differences can be seen only when looking at the total amplitude
$r|\Psi_4|_{22}/M_{\odot}$ (see the inset).

\begin{figure}[t]
\begin{center}
   \includegraphics[width=0.49\textwidth]{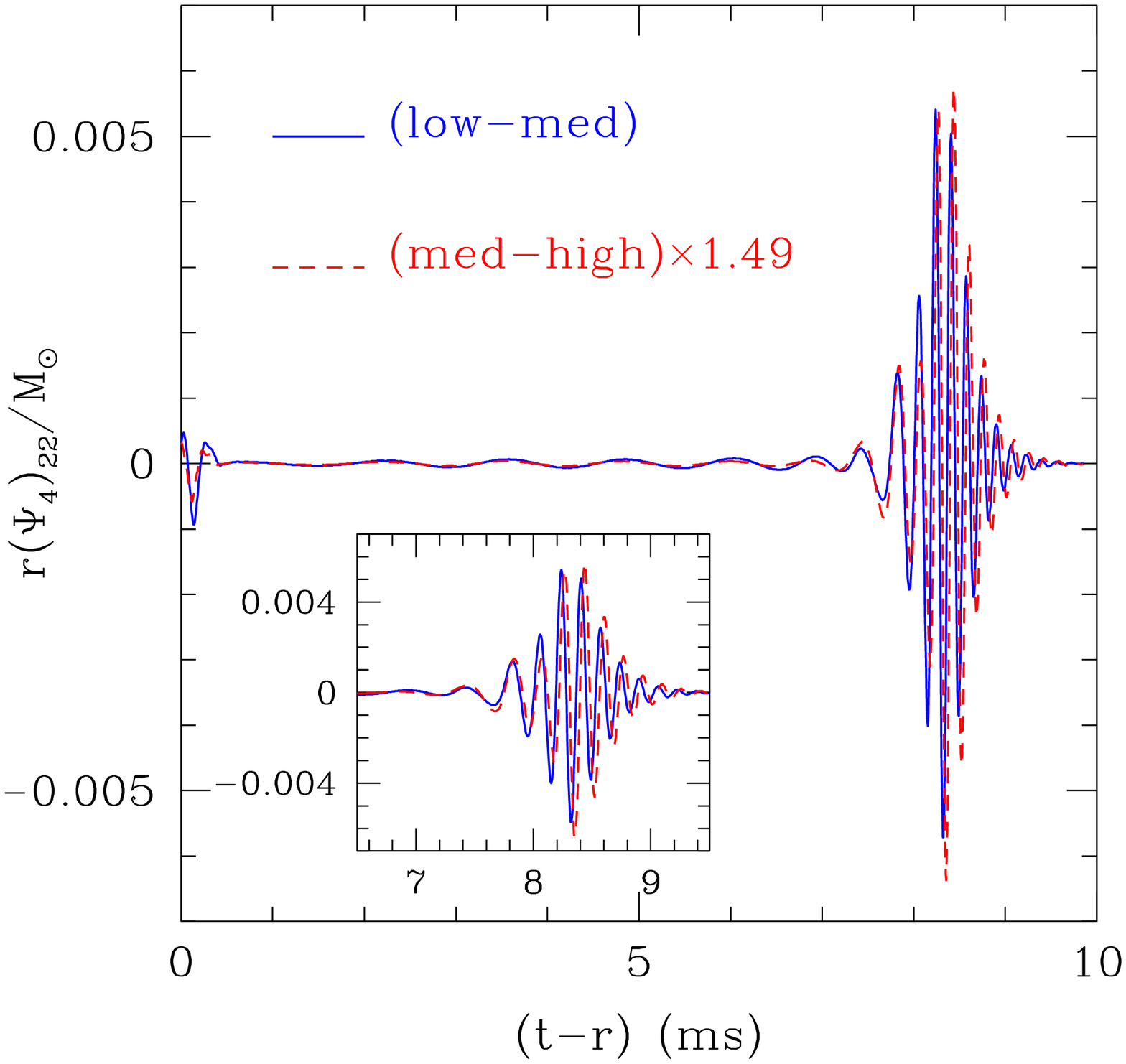}
   \includegraphics[width=0.49\textwidth]{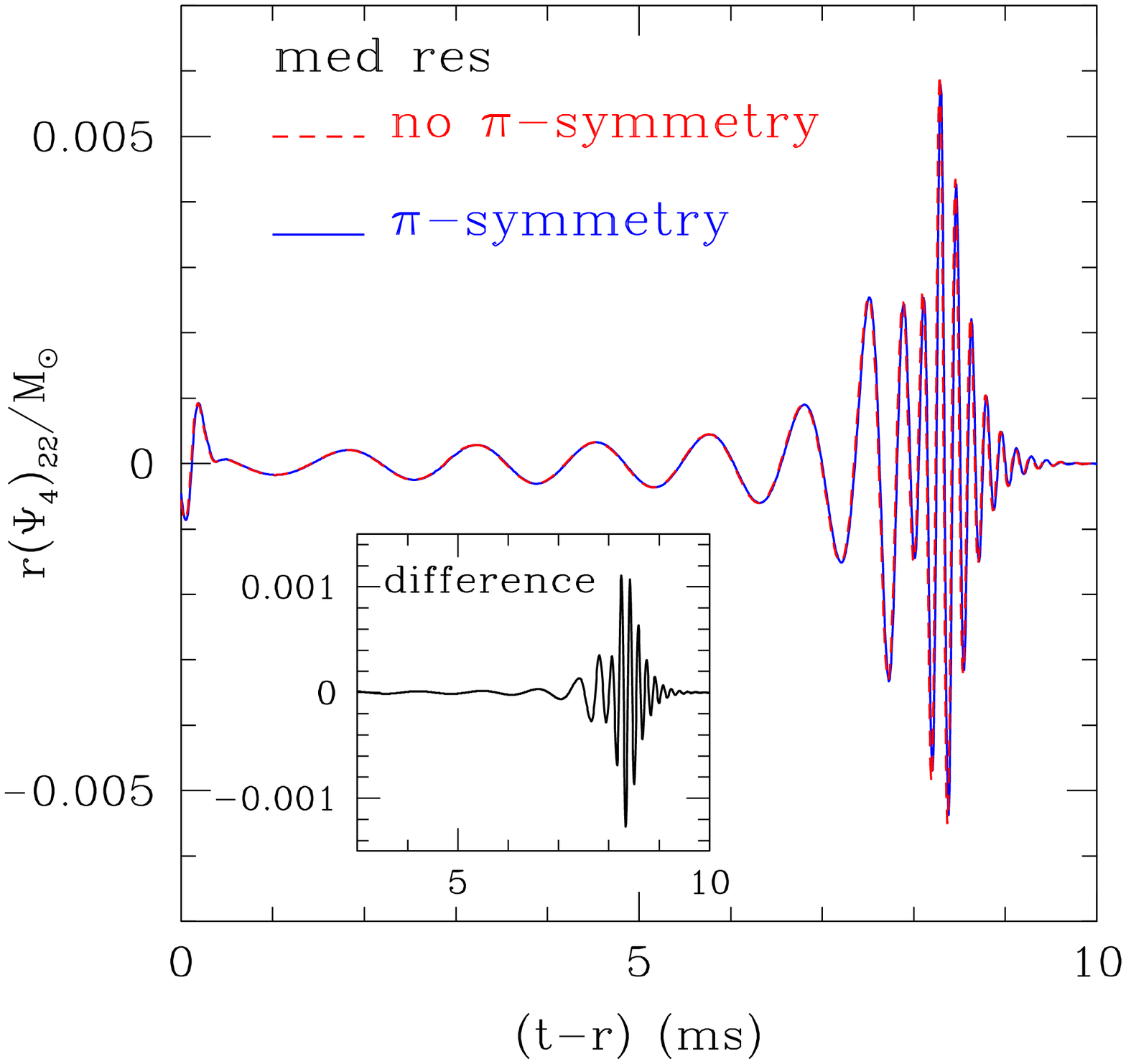}
\end{center}
\vskip -0.5cm
   \caption{\textit{Left panel:} Convergence test for
     $r(\Psi_4)_{22}/M_{\odot}$ waveform as extracted at
     $r=200\,M_{\odot}$. Different lines show the residuals between
     the waveforms computed at different resolutions, suitably shifted
     in time and scaled to compensate for a convergence rate of
     $1.8$. \textit{Right panel:} Waveforms computed either enforcing
     a $\pi-$symmetry (blue continuous line) or not (red dashed line).
     Indicated in the inset is the difference between the two
     waveforms.}
     \label{fig:waves_conv}
\end{figure}

In the left panel of Figure~\ref{fig:waves_conv} we present a
convergence test for the $\Psi_4$ waveform as extracted at
$r=200\,M_{\odot}$, where different lines represent the differences
between the waveforms suitably shifted in time and scaled to
compensate for the different truncation errors. As discussed in detail
in~\cite{Pollney:2007ss}, the dominant source of error is a de-phasing
which causes the lower resolution evolutions to ``lag'' behind the
higher resolution. This delay is usually rather small and between
$\sim 2\times 10^{-2}$ and $\sim 5\times 10^{-2}\,\msec$, but it is
clearly visible when comparing the total amplitude of $\Psi_4$ as a
function of time. If not properly taken into account, this error
spoils the convergence tests, making the residuals appear as over or
under convergent (\cf Figure 2 of ~\cite{Pollney:2007ss}). This is
obviously an artifact of the near cancellation of the lowest-order
terms in the truncation error and it is induced by the small
time-differences at different resolutions. We have removed this
contamination by shifting the time coordinate of the low and medium
resolution runs by the time interval needed to produce an alignment of
the maxima of the emitted radiation (details on how to do this are
discussed in Appendix A of~\cite{Pollney:2007ss}). Once this
correction is made, the rescaled residual errors overlap reasonably
well, indicating a convergence rate of $1.8$; this is most likely due
to the fact that the evolution is $2$nd-convergent for most of the
time and because the coefficients of the $\sim {\cal O}(h)$ truncation
error are much smaller than the corresponding $\sim {\cal O}(h^{2})$
one. To the best of our knowledge this is the first time that such a
convergence test is shown on the gravitational waves from neutron-star
binaries.

Finally in the right panel of Figure~\ref{fig:waves_conv} we report
a comparison of waveforms computed either enforcing a $\pi-$symmetry (blue
continuous line) or not (red dashed line). Clearly the two are
essentially indistinguishable on this scale and the small phase differences are
shown in the inset.

\section{Variations on the Theme and ``Error Budget''}
\label{sec:errors}

In the previous Sections we have discussed in detail the accuracy of
our simulations for a fiducial high-mass binary by analysing how the
results change with resolution. However, the resolution is not the
only aspect of these simulations which can be modified to yield
slightly different results, although it is unquestionably one of the most
important ones. This Section is therefore dedicated to those purely
numerical aspects which can be varied and have been seen to yield
(small) changes in the results. This assessment is obviously
computationally very expensive, but it has allowed us to compile an
\textit{``error budget''} providing a simple reference on the
influence that different sources of inaccuracies have on the physical
results.

We build this budget by considering the evolution of the modulus of
the waveform $|(\Psi_4)_{22}|$ (see, \eg Figure 20 of paper I or the
inset in the right panel of Figure~\ref{fig:waves}) and by singling
out two specific times corresponding to the first and last
maximum\footnote{When the collapse to black hole is delayed, the HMNS
  emits large amounts of gravitational radiation, thus leading to a
  series of local maxima in the evolution of $|(\Psi_4)_{22}|$ (see,
  \eg Figure 26 of paper I)}; we define these two times as the ``time
of merger'', $t_{\rm merg.}$, and the ``time of collapse'', $t_{\rm
  coll.}$. For the high-resolution simulation described
here, these are $t_{\rm merg.}=8.7\,\msec$ and $t_{\rm
  coll.}=9.3\,\msec$, respectively. Note that the merger time defined
above should not be confused with the time of the ``hydrodynamical
merger'', namely the time when the two stellar cores merge, $t_{\rm
  hd-merg.}$, and which takes place considerably earlier ($t_{\rm
  hd-merg.}=7.4\,\msec$ for the high-resolution case); we mark this
hydrodynamical merger as the time when the maximum rest-mass density
has a first significant minimum (\cf~Figures~2 or 8
in~\cite{Baiotti08}).

\begin{table*}[t]
  \caption{``Error-budget'', \ie variation of physical quantities
    induced by changes in the numerical settings. The table reports:
    the numerical setting varied, the amount of such a variation with
    respect to our standard simulation, and the percentile change of
    the time of the merger, $t_{\rm merg.}$, and of that of the
    collapse, $t_{\rm coll.}$ (see text for definitions), with respect
    to our standard simulation. A $\pm$ sign indicates whether the
    time is increased ($+$) or decreased ($-$) with respect to the
    fiducial medium-resolution run.  }
\vskip 0.25cm
\begin{tabular}{l|l|c|c}
Numerical setting varied & variation & $\Delta t_{\rm  merg.}\ \ (\%)$& $\Delta t_{\rm coll.}\ \ (\%)$\\
\hline
resolution   & $\times\ 1.25$ (from medium to high)      & $+0.45$ & $+0.37$ \\
extent of finest grid  & $\times\ 4$            & $+0.6$ & $+0.6$ \\
spatial prolongation order        & $3$rd vs. $5$th order  & $+0.2$ & $+0.3$ \\
artificial dissip. coeff.&$\times\ 100$ (from $0.1$ to $0.001$)& $+0.1$ & $+0.1$ \\
outer-boundary location           & $\times\ 2$            & $-0.06$& $+0.1$ \\
\end{tabular}
\\
\label{table:error}
\end{table*}

The error budget is reported in Table~\ref{table:error}, whose rows
contain information about the specific numerical setup which has been
varied, the amount of the variation and the differences in the two
times. Note that because the HMNS collapses promptly to a black hole,
the differences in $t_{\rm coll.}$ reported here are particularly
small and should therefore be taken as lower limits. Indeed, for less
massive HMNSs, the corresponding differences can be as large as $\sim
10\%$ and increase to $\gtrsim 70\%$ when considering simulations run
at low resolution. The determination of $t_{\rm coll.}$ represents
therefore an emblematic example of how physically reasonable and
numerically convergent results obtained at low-resolution suffer from
very large truncation errors.

The first row in Table~\ref{table:error} reports the variations due to an
increase in the resolution (\ie from the medium resolution to the high
one) and does not need further comment besides noting that the results
for the subsequent rows refer to medium-resolution simulations. The
second row shows the change induced when the extent of the finest grid
is increased by a factor of $4$ in all the three spatial
directions. In this case, a single fine grid covers both stars, so
that the volume of the finest grid is $4^3$ times larger and the
simulation is therefore about $64$ times more expensive. The third row
reports the variations measured when the interpolation for the spatial
prolongation operation needed in the time-stepping of our mesh
refinements is changed from $3$rd to $5$th-order (only for the
spacetime variables). The fourth row refers to a variation in the
coefficient of the artificial-dissipation (see discussion
in~\cite{Baiotti06,Baiotti07}). The value of the dissipation
coefficient $\varepsilon$ in the fiducial simulation is $0.1$ and the
table reports the changes when using instead $\varepsilon=0.001$;
clearly the differences are minute but a higher dissipation also
provides waveforms which are less affected by reflections among
different refinement levels. Finally, the fifth row reports the
changes when the outer-boundary location is increased by a factor
$2$. While the content of the table is self-explanatory, it is worth
remarking that in all cases the induced variations are below $1\%$ at
these resolutions and that the times are essentially always increased.


\section{Conclusions}
\label{sec:conclusion}

We have presented a detailed analysis of the accuracy and convergence
properties of our general-relativistic simulations of the inspiral and
merger of binary neutron stars. More specifically, we have shown that
for a high-mass binary and typical resolutions of $h\sim
0.19\,M_{\odot}- 0.12\,M_{\odot}$, the results show the expected
convergence rate of $1.8$ during the inspiral phase, the collapse and
the subsequent ringdown.  However, the convergence rate drops to $1.2$
at the merger and during the evolution of the HMNS. This deterioration
of the convergence rate is not surprising and can be explained as a
combination of (at least) two effects. Firstly, strong shocks form
during the merger and HRSC schemes are only $1$st-order accurate at
discontinuities. Secondly, the merger gives rise to turbulent
motions, related to the development of a Kelvin-Helmholtz instability,
for which the concept of local convergence needs to be revisited.

We have also presented tests on the conservation of physical
quantities, such as the rest mass, the energy and the angular
momentum, taking into account that the last two are not conserved
because of losses in gravitational-wave. In particular, we have shown
that, at our best resolution, the rest-mass is conserved with an
error $\lesssim 10^{-6}$, while the energy and the angular momentum
are conserved to $\lesssim 1\%$ after taking into account the parts
lost to radiation. In addition, when considering the accuracy of the
extraction of gravitational-wave information, we have shown the very
good agreement, in both phase and amplitude, of the gravitational
waves extracted from different detectors within the same simulation,
or from the same detector but at different resolutions. Such waveforms
have been shown to be also convergent at a rate of $1.8$. Finally, we
have reported a first investigation on how purely numerical changes in
some of the setting of the simulation can influence the physical
results. This analysis has lead to the construction of a
straightforward ``error budget'', whose entries are all below $1\%$,
at least for the high-mass binary considered here.

A final remark should be made when contrasting binary-black-hole
simulations with the corresponding ones involving binary neutron
stars.  While for the first ones the numerical methods are
sufficiently robust (and the convergence order sufficiently high) that
an increase in resolution is usually the solution to the most serious
problems, the hydrodynamical complexities inherent to the second ones
(and the small convergence order after the merger) are such to require
extra caution and very careful assessment of the possible sources of
error. Experience has shown us that results at low resolution which
appear reasonable and convergent, are however contaminated by large
truncation errors and hence of little physical relevance. In view of
this, and of the low convergence rate after the merger, we conclude
with a remark that obvious in general, but is worth making when
considering binary neutron star calculations: high-order methods and
the highest possible resolutions are imperative to draw robust
conclusions.

\ack 
	We thank the developers of \texttt{Lorene} for providing us
        with initial data, those of \texttt{Carpet} for the mesh
        refinement, and those of \texttt{Cactus} for the computational
        infrastructure our code is based on. Useful comments from
        Carlos Palenzuela and Masaru Shibata are kindly
        acknowledged. The computations were performed at the AEI and
        at LONI (\texttt{www.loni.org}). This work is also supported
        by the DFG SFB/Transregio~7 and by the JSPS grant 19-07803.

\section*{References}
\bibliographystyle{iopart-num}
\bibliography{aeireferences}

\end{document}